\documentclass[preprintnumbers,amsmath,11pt,amssymb,floatfix,superscriptaddress,
showpagenumber,showpacs,nofootinbib]{revtex4}
\usepackage{graphicx}
\usepackage{epsfig}
\usepackage{bm}
\usepackage{amsfonts}

\input epsf

\begin{document}

\title{\Large Israel-Stewart approach to viscous dissipative extended holographic Ricci dark energy dominated universe}

\author{Surajit Chattopadhyay}
\email{surajitchatto@outlook.com, surajcha@associates.iucaa.in}
\affiliation{ Pailan College of Management and Technology (MCA Division), Bengal
Pailan Park, Kolkata-700 104, India.}

\date{\today}

\begin{abstract}
This paper reports a study on the truncated Israel-Stewart formalism for bulk viscosity using the extended holographic Ricci dark energy (EHRDE), where the
density of dark energy is given as a combination of the Hubble parameter and its derivative. The equations of motion are integrated and the resulting model
is analysed under many aspects, including the state finder parameters and the generalised thermodynamics second law. Under the consideration that the universe is dominated by EHRDE the evolution equation for the bulk viscous pressure $\Pi$ in the framework of the truncated Israel-Stewart theory has been taken as $\tau \dot{\Pi}+\Pi=-3\xi H$, where $\tau$ is the relaxation time and $\xi$ is the bulk viscosity coefficient. Considering effective pressure as a sum of thermodynamic pressure of EHRDE and bulk viscous pressure it has been observed that under the influence of bulk viscosity the EoS parameter $w_{DE}$ is behaving like phantom i.e. $w_{DE}\leq -1$. It has been observed that the magnitude of the effective pressure $p_{eff}=p+\Pi$ is decaying with time. We also investigated the case for a specific choice of scale factor namely $a(t)=(t-t_0)^{\frac{\beta}{1-\alpha}}$. For this choice we have observed that a transition from quintessence to phantom is possible for the equation of state parameter. However, the $\Lambda$CDM phase is not attainable by the statefinder trajectories for this choice. Finally it has been observed that in both of the cases the generalized second law of thermodynamics is valid for the viscous EHRDE dominated universe enveloped by the apparent horizon.
\end{abstract}

\maketitle

\maketitle
\section{Introduction}
Accelerated expansion of the current universe was reported by Riess et al. \cite{S27} of high-redshift supernavoe search team and Perlmutter et al. \cite{S271} of supernovae cosmology project team through accumulation of observational data from distant type Ia Supernovae. Discovery of
\cite{S27,S271} was truly ground breaking and subsequently this has been further confirmed by other observational studies including more detailed studies of
supernovae and independent evidence from clusters of galaxies, large-scale structure (LSS) and the cosmic microwave background (CMB)\cite{Fried}. The reason
behind this expansion is referred to as ``dark energy" (DE), which is regarded as an exotic matter characterized by negative pressure and having equation of state parameter $w=p/\rho<-1/3$ required for accelerated expansion of the universe. Nature of this DE is not yet clear and different DE candidates have been proposed till date. Some remarkable reviews on DE include \cite{copeland-2006,bambareview,DErev,nojirireview}. While proposing a phantom cosmology based unifying approach to early and late-time universe Nojiri and Odintsov \cite{nojiriadd} suggested generalized holographic dark energy (HDE)involving infrared cut-off combined with FRW parameters and also discussed the entropy bound in phantom era. HDE is based on ``holographic principle"  and the density of HDE is \cite{3}
$\rho_{\Lambda}=3c^2M_p^2L^{-2}$, where $L$ is the infrared cut-off. Some notable works on HDE include \cite{holo1,holo2,holo3,holo4,holo51,holo61}. Furthermore, there exist plethora of literatures on HDE in theoretical aspects as well as observational constraints e.g. \cite{holo5,holo6,holo7}. 

In the present work we consider
a special form of HDE \cite{EHRDE1} dubbed as ``extended
holographic Ricci dark energy" (EHRDE) \cite{EHRDE2,EHRDE3}, whose
density has the form
\begin{equation}\label{RD}
\rho_{DE}=3 M_p^2\left(\alpha H^2+\beta \dot{H}\right)
\end{equation}
where the upper dot represents derivative with respect to cosmic time $t$, $M_p^2$ is the reduced Planck mass, $\alpha$ and $\beta$
are constants to be determined. In this context we should mention that the role of a distance proportional to the Ricci scale as a causal connection scale for perturbations was noticed in \cite{R1} and \cite{R2} used it for the first time as a DE cutoff scale. Detailed cosmology of Ricci DE was discussed in \cite{R3} and some further literatures on Ricci DE include \cite{R4,R5,R6,R7}. Wang and Xu
\cite{2} found the best-fit values in order to make the cutoff of \cite{EHRDE2} to
be consistent with observational data as
$\alpha=0.8502^{+0.0984+0.1299}_{-0.0875-0.1064}$ and
$\beta=0.4817^{+0.0842+0.1176}_{-0.0773-0.0955}$. In the current work we shall take
$\alpha= 0.98$ and $\beta=0.37$.
In the early evolution of the universe, dissipative effects, including both bulk and shear viscosity, are supposed to play a very important role \cite{vcg}. Chimento et al \cite{vis1} had shown that accelerated expansion can be derived by the combination of a cosmic fluid with bulk dissipative pressure and quintessence matter and it can also solve the coincidence problem. Evolution of the universe involves a sequence of important dissipative processes that includes GUT phase transition at $t\approx 10^{-34} s$ and a temperature of about $T\approx 10^{27}K$ \cite{vcg}.
Eckart \cite{eckart} and Landau and Lifshitz \cite{landu} were the first ones to attempt at creating a theory of relativistic dissipative fluids. A relativistic second-order theory was developed by Israel and Stewart \cite{israel}.

 A time dependent viscosity consideration was made to DE by Nojiri and Odintsov \cite{nojiriadd2} while considering EoS with inhomogeneous, Hubble parameter dependent term and it was demonstrated that the thermodynamic entropy may be positive even in phantom era as a result of crossing the phantom boundary. Pun et al \cite{vcg} considered a generalization of the
Chaplygin gas model and by assuming the presence of a bulk viscous type dissipative term in the effective
thermodynamic pressure of the gas and by considering dissipative effects described by truncated Israel-Stewart model they \cite{vcg} had shown that viscous Chaplygin gas model offers an effective dynamical possibility for replacing the cosmological constant. Effect of viscosity in Chaplygin gas with different modifications have been studied in \cite{vcg1,vcg2,vcg3}. Cataldo et al \cite{vis2} investigated dissipative processes in the universe in the full causal Israel-Stewart-Hiscock theory and showed that the negative pressure generated by the bulk viscosity cannot avoid that the dark energy of the universe to be phantom energy. Brevik and Gorbunova \cite{vis3} assumed the bulk viscosity to be proportional to the scalar expansion in a spatially flat FRW universe and had shown that it could lead the universe to phantom phase even if the universe lies in the quintessence phase in the non-viscous scenario. Ren and Meng \cite{vis4} considered a generally parameterized EoS in the cosmological evolution with bulk viscosity media modelled as dark fluid. Setare and Sheykhi \cite{setare2} studied the validity of generalized second law of thermodynamics in presence of viscous dark energy in a non-flat universe and concluded with validity of the generalized second law. Feng and Li \cite{feng} investigated viscous Ricci dark energy and showed that once viscosity is taken into account the problem on age of the universe gets alleviated. Amirhashchi and Pradhan \cite{vis5} considered viscous and non-viscous dark energy EoS parameter in anisotropic Bianchi type I space-time and could get a transition from quintessence to phantom. Brevik et al. \cite{vis6} investigated interacting dark energy and dark matter in flat FRW universe taking bulk viscosity as a function of Hubble parameter $H$ and cosmic time $t$ and studied subsequent corrections of thermodynamical parameters. Velten et al.\cite{vis7} derived conditions under which either viscous matter or radiation cosmologies can be mapped into the phantom dark energy scenario with constraints from multiple observational data sets. Jamil and Farooq \cite{vis8} presented a generalization of interacting HDE using the viscous generalized Chaplygin gas and reconstructed the potential and the dynamics of the scalar field. Setare and Kamali \cite{vis9} study warm-viscous inflationary universe model on the brane in a tachyon field theory and obtained the general conditions which are required for this model to be realizable. In a very recent work, Bamba and Odintsov \cite{Bamba1} investigated a fluid model in which
EoS for a fluid includes bulk viscosity and found that the spectral index of the curvature perturbations, the tensor-to-scalar ratio of the density perturbations, and the running of the spectral index, can be consistent with the recent Planck results. It may be noted that in a remarkable work Brevik et al. \cite{nojiriadd3} discussed entropy of DE filled FRW universe in the framework of holographic Cardy-Verlinde formula and expressed entropy in terms of energy and Casimir energy depending on the EoS and in a relatively recent work, Brevik et al. \cite{ nojiriadd4} derived a formula for the entropy for a multicomponent coupled fluid that under certain conditions may reduce to the Cardy-Verlinde form to relate the entropy of a closed FRW universe to the energy contained in it together with its Casimir energy. In some recent studies bulk viscosity has been incorporated in the studies of modified gravity too and studies in this direction include \cite{mod1,mod2,mod3}.

Plan of the present work is as follows: In section II we shall apply Israel-Stewart theory to study the behavior of viscous extended holographic Ricci dark energy (EHRDE) without any specific choice of scale factor as well as for the choice $a(t)=(t-t_0)^{\frac{\beta}{1-\alpha}}$. In section III we shall study the statefinder parameters and investigate whether $\Lambda$CDM phase is attainable for both the cases. In section IV we shall examine validity of the generalized second law of thermodynamics for both the cases and we in section V we shall conclude.

\section{Israel-Stewart approach }
\subsection{Israel-Stewart approach without any specific choice of scale factor}
In the cosmological framework bulk viscosity can be thought of as an internal friction due to
the different cooling rates in an expanding gas \cite{vcg}. As the dissipation due to bulk viscosity converts kinetic energy of the
particles into heat, the effective pressure is expected to be reduced in an expanding fluid. For a flat homogeneous Friedmann-Robertson-Walker (FRW) with a line element:
\begin{equation}
ds^2=dt^2-a^2(t)\left(dx^2+dy^2+dz^2\right)
\end{equation}
filled with a bulk viscous cosmological fluid the energy-momentum tensor is given by \cite{vcg}
\begin{equation}
T_i^k=(\rho+p+\Pi)u_iu^k-(p+\Pi)\delta_i^k
\end{equation}
where $\rho$, $p$ and $\Pi$ are energy density, thermodynamic pressure and the bulk viscous pressure respectively. The $u_i$ is four velocity
that satisfies the condition $u_iu^i=1$. Here, $N^i=nu^i$ and $S^i=\sigma N^i-\left(\tau \Pi^2/2\xi T\right)u^i$ are particle and entropy fluxes respectively, where $n$, $\sigma$ and $T\geq 0$ implies the number density, the specific entropy, and temperature respectively. Also, bulk viscosity coefficient and relaxation time are $\xi$ and $\tau\geq 0$ respectively. If the Hubble parameter is $H=\frac{\dot{a}}{a}$ then the gravitational field equations together with the continuity equation are \cite{vcg}
\begin{eqnarray}
  3H^2 &=& \rho, \label{f1}\\
  2\dot{H}+3H^2 &=& -p-\Pi, \label{f2}~~~\\
\dot{\rho}+3H(\rho+p) &=&- 3H\Pi. \label{f3}~~~~~~~
\end{eqnarray}
The effect of the bulk viscosity can be considered by
adding thermodynamic pressure $p$ to the bulk viscous pressure $\Pi$, i.e.
\begin{equation}
p_{eff}=p+\Pi
\end{equation}
terms in the energy-momentum tensor
Taking $x=\ln a$ in Eq. (\ref{RD}) we have ($M_p^2=1$)
\begin{equation}\label{RDx}
\rho_{DE}=3 \left(\alpha H^2+\frac{\beta}{2} \frac{dH^2}{dx}\right)
\end{equation}
 Considering $\rho=\rho_{DE}$ in Eq.(\ref{f1}) we have the Hubble parameter
\begin{equation}\label{Hx}
H(x)=H_0 \exp \left(\frac{x(1-\alpha)}{ \beta}\right)
\end{equation}
Putting $x=\ln a$ in Eq. (\ref{Hx}) we can write
\begin{eqnarray}
H=H_0 a^{\frac{1-\alpha}{\beta}}
\end{eqnarray}
which can be written in the form of a differential equation
\begin{eqnarray}
\dot{a}(t)=H_0 a(t)^{1+\frac{1-\alpha}{\beta}}
\end{eqnarray}
where the upper dot implies time derivative with respect to cosmic time $t$. Since for the current time $t=t_0$ we have $a=1$, we have the following particular solution for $a(t)$:
\begin{equation}\label{a}
a(t)=\left(\frac{H_0 (t-t_0) (-1+\alpha )+\beta }{\beta }\right)^{\frac{\beta }{-1+\alpha }}
\end{equation}
Hence, Hubble parameter as expressed in Eq. (\ref{Hx}) can be written as a function of $t$ :
\begin{equation}\label{Ht}
H=\frac{H_0 \beta}{H_0 (t-t_0)(-1+\alpha)+\beta}
\end{equation}
Using Eq. (\ref{Ht}) in (\ref{RD}) we have the reconstructed RDE as
\begin{equation}\label{RDreconst}
\rho_{DE}=3H^2=\frac{3 H_0^2  \beta ^2}{\left(H_0 (t-t_0) \left(-1+ \alpha \right)+ \beta \right)^2}
\end{equation}
Using Eq. (\ref{RDreconst}) in (\ref{f3}) we have the EoS parameter
\begin{equation}\label{EoS}
w_{DE}=-1+\frac{1}{3} \left(-\frac{2}{ \beta }+\frac{2 \alpha }{\beta }-\frac{\Pi \left(H_0 (t-t_0) \left(-1+
\alpha \right)+ \beta \right)^2}{H_0^2  \beta ^2}\right)
\end{equation}
Hence, for present acceleration i.e. $w<-1/3$ at $t=t_0$ we need
\begin{equation}
\alpha<1+\frac{\beta}{2}\left(6+\frac{\Pi(t=t_0)}{H_0^2}\right)
\end{equation}
If the current universe is in phantom phase i.e. $w<-1$ then we shall require
\begin{equation}
\alpha<1+\frac{\Pi(t=t_0)}{2H_0^2}\beta
\end{equation}
For the bulk viscosity coefficient $\xi$ and for the relaxation time $\tau$ of the viscous extended holographic Ricci dark energy we assume the following
phenomenological laws \cite{vcg}
\begin{equation}\label{xi}
\xi=\eta \rho^{\nu},~~~~~~~~~~~~\tau=\xi \rho^{-1}=\eta\rho^{\nu-1}
\end{equation}
where, $\eta\geq 0$ and $\nu\geq 0$ are constants. At this juncture it may be stated that the bulk viscosity
coefficient $\xi$ is being considered as a function of $\rho(t)$. Hence possibility is open to a variety of
$\xi(\rho(t))$. The case $\nu=1/2$ yields $\xi=\eta \rho^{1/2}$ that corresponds to a power-law expansion for the
scale factor. To obtain solution with big-rip no restriction is imposed on $\nu$ and similar approach
was adopted in \cite{vis2}. Another work that is noteworthy in this context is done by Colistete et al.
\cite{Colistete}, where for viscous generalized Chaplygin gas a variety of solutions were presented for different
ranges of $\nu$. Taking $\nu=2$ and subsequently solving the evolution equation for $\Pi$ in the framework of the truncated Israel-Stewart theory given by
\begin{equation}\label{Pi}
\tau \dot{\Pi}+\Pi=-3H\xi
\end{equation}
we obtain the evolution equation of the bulk viscous pressure as
\begin{equation}\label{Pisolve}
\Pi=e^{-\frac{(\varphi_1 t+\varphi_2)^3}{9 \varphi_2 H_0^2 \eta }} C_1-\frac{e^{-\frac{(\varphi_1 t+\varphi_2)^3}{9 \varphi_2 H_0^2 \eta }} H_0^3 \left(-9 e^{\frac{(\varphi_1 t+\varphi_2)^3}{9 \varphi_2 H_0^2 \eta }}+3^{2/3} \left(-\frac{(\varphi_1 t+\varphi_2)^3}{\varphi_2 H_0^2 \eta }\right)^{2/3} \Gamma\left[\frac{1}{3},-\frac{(\varphi_1 t+\varphi_2)^3}{9
\varphi_2 H_0^2 \eta }\right]\right)}{2\varphi_2 (\varphi_1 t+\varphi_2)^2}
\end{equation}
where
\begin{eqnarray}
  \varphi _1=\frac{H_0}{\beta } ( \alpha -1 ),\nonumber \\
  \varphi _2=1+\frac{H_0 t_0}{\beta }\left(1- \alpha \right). \nonumber
\end{eqnarray}
and $C_1$ is the constant of integration.
Hence, EoS parameter, when expressed as a function of $x(=\ln a)$ takes the form
\begin{equation}\label{wDE}
\begin{array}{c}
w_{DE}=-1+\frac{1}{3} \left(\frac{2 (\alpha -1)}{\beta }-\frac{1}{2 H_0^2}e^{\frac{2 x (\alpha-1 )}{\beta }-X} \left(2-\frac{H_0^3
\beta  \left(-9 C_1 e^X+3^{2/3} (-X)^{2/3} \Gamma\left[\frac{1}{3},-X\right]\right)}{(H_0 (t_0(1-\alpha )+\beta )) \left(\frac{X \left(9
H_0^2 (H_0 t_0(1-\alpha )+\beta ) \eta \right)}{\beta }\right)^{\frac{2}{3}}}\right)\right)
\end{array}
\end{equation}
where
\begin{equation}\label{X}
X=\frac{\beta  \left(\frac{H_0 (\alpha-1 )}{\beta }+\left(1-\frac{H_0 t_0 (\alpha-1 )}{\beta }\right) \left(t_0+\frac{\left(-1+e^{\frac{x
(\alpha-1 )}{\beta }}\right) \beta }{H_0 (\alpha-1 )}\right)\right)^3}{H_0^2 (H_0 t_0(1-\alpha) +\beta ) \eta
}
\end{equation}
Eq.(\ref{wDE}) imposes one more constraint on the relationship between $\alpha$ and $\beta$ as
\begin{equation}
\alpha\neq 1+\frac{\beta}{H_0t_0}
\end{equation}
We would like to add a note at this juncture. Eq. (\ref{xi}) comes as a phenomenological law for bulk
viscosity coefficient and relaxation time with $\eta\geq 0$ i.e. possibility of $0$ is not excluded. Clearly,
for $\eta=0$ the bulk viscous pressure will vanish and we shall get back the non-viscous scenario.
However, it is clear that the evolution equation for $\Pi$ in the framework of Israel-Stewart theory
is written for viscous scenario and for non-viscous it is irrelevant. Eq.(\ref{Pi}) is a linear differential
equation on $\Pi$ whose integrating factor is $\exp\int\frac{dt}{\tau}$ and obviously the integrating factor will not
exist if $\eta=0$ and hence there will exist no solution and Eq.(\ref{X}) will be of no existence.

From Eq.(\ref{wDE}) we observe that for $\alpha=0.98,~\beta=0.37$ in EHRDE as mentioned in the previous section we have for very late stage of the universe
\begin{equation}
\begin{array}{c}
x\rightarrow \infty \Rightarrow w_{DE}\rightarrow -1+
\frac{1}{3} \\
\left(-0.108-\frac{1}{H_0^2}-\left.\left(10^{-5} H_0^3 \left(-9+19577.259 \left(\frac{\left(0.137+0.0148 H_0 t_0+
4\times
10^{-4} H_0^2 \left(-1+t_0^2\right)\right)^3}{H_0^5 (-0.37-0.02 H_0 t_0) \eta }\right)^{2/3}\right.\right.\right.\right. \\ \left.\left.\left.\left.\Gamma\left[\frac{1}{3},\frac{101452.804
\left(0.1369+0.0148 H_0 t_0+4\times 10^{-4}H_0^2 \left(t_0^2-1\right)\right)^3}{H_0^5 (-0.37-0.02 H_0 t_0)
\eta }\right]\right)\right)\right/\right.\\
\left.\left((-0.37-0.02H_0 t_0) \left(0.1369+0.0148 H_0 t_0+4\times 10^{-4} H_0^2 \left(t_0^2-1\right)\right)^2\right)\right)
\end{array}
\end{equation}
and for early stage
\begin{equation}
\begin{array}{c}
x\rightarrow -0.14 \Rightarrow w_{DE}\rightarrow-1+
\frac{1}{3}
\left(-\frac{1}{H_0^2}0.5e^{-\frac{0.04 \left(-0.14+0.99 H_0 t_0+H_0^2 \left(-0.054+0.054 t_0^2\right)\right)^3}{H_0^5
(0.37+0.02 H_0 t_0) \eta }} \right.\\
\left(2-\left.\left(0.37H_0^3 \left(-9e^{\frac{0.04 \left(-0.14+0.99 H_0 t_0+H_0^2 \left(-0.054+0.054 t_0^2\right)\right)^3}{H_0^5
(0.37+0.02 H_0 t_0) \eta }}+\right.\right.\right.\right.\\
\left.\left.\left.\left.1.07 \left(-\frac{ \left(-0.14+0.99 H_0 t_0+H_0^2 \left(-0.054+0.054t_0^2\right)\right)^3}{H_0^5
(0.37+0.02 H_0 t_0) \eta }\right)^{2/3} \Gamma\left[0.33,-\frac{0.04 \left(-0.054 H_0+\left(-\frac{0.14}{H_0}+t_0\right)
(1+0.054 H_0 t_0)\right)^3}{H_0^2 (0.37+0.02 H_0 t_0) \eta }\right]\right)\right)\right/\right.\\
\left.\left.\left((0.37+0.02 H_0 t_0) \left(-0.054 H_0+\left(-\frac{0.14}{H_0}+t_0\right) (1+0.054 H_0 t_0)\right)^2\right)\right)\right)
\end{array}
\end{equation}
where $\Gamma(a,z)=\int_{z}^{\infty}e^{-t}t^{a-1}dt$.

From the above limits one can see that the EoS parameter is determined by the viscosity of the EHRDE and does not blow up in the past or future. This observation is consistent with \cite{feng}, where the RDE was considered with barotropic fluid and the EoS parameter was seen not to blow up in the past of the future.
\begin{figure}[h]
\includegraphics[width=20pc]{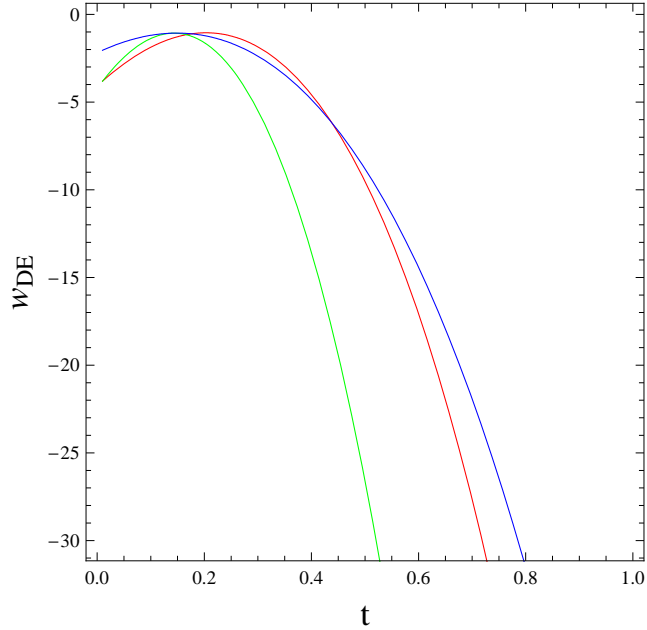}
\caption{\label{figEoS}EoS parameter based on Eq.(\ref{wDE}). Red, green and blue lines correspond to $\{\alpha,\beta\}$ combination of $\{0.9735,0.3701\}$ with $\eta=3.5\times 10^{-4},3.0\times 10^{-4}$ and $2.5\times 10^{-4}$ respectively.  }
\end{figure}
\begin{figure}[h]
\includegraphics[width=20pc]{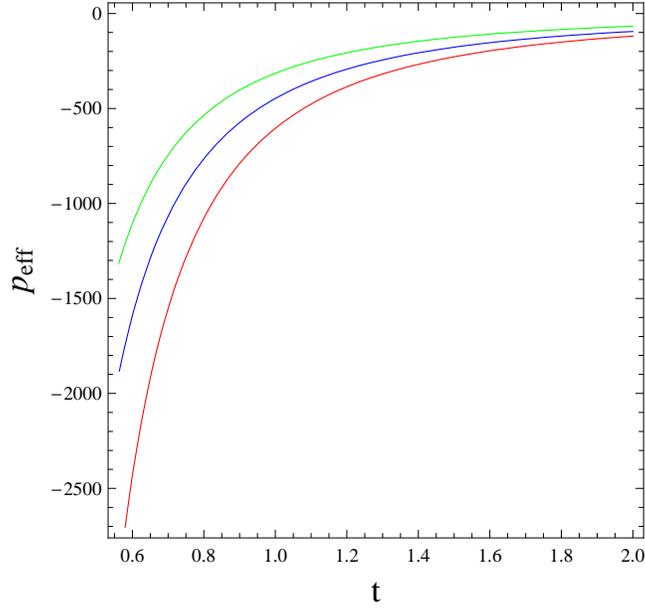}
\caption{\label{figpeff}Plot of $p_{eff}=p+\Pi$. Red, green and blue lines correspond to $\{\alpha,\beta\}$ combination of $\{0.9735,3701\}$ and $\{0.9735,3701\}$ with $\eta=3.5\times 10^{-4},3.0\times 10^{-4}$ and $2.5\times 10^{-4}$ respectively.}
\end{figure}
\begin{figure}[h]
\includegraphics[width=20pc]{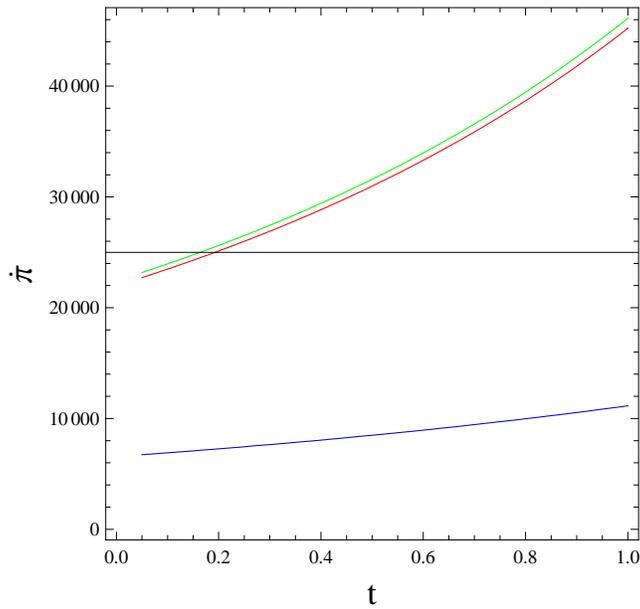}
\caption{\label{figpidot}Plot of $\dot{\Pi}$ based on Eq.(\ref{Pisolve}). Red, green and blue lines correspond to $\{\alpha,\beta\}$ combination of $\{0.9735,3701\}$ with $\eta=3.5\times 10^{-4},3.0\times 10^{-4}$ and $2.5\times 10^{-4}$ respectively.}
\end{figure}
\begin{figure}[h]
\includegraphics[width=20pc]{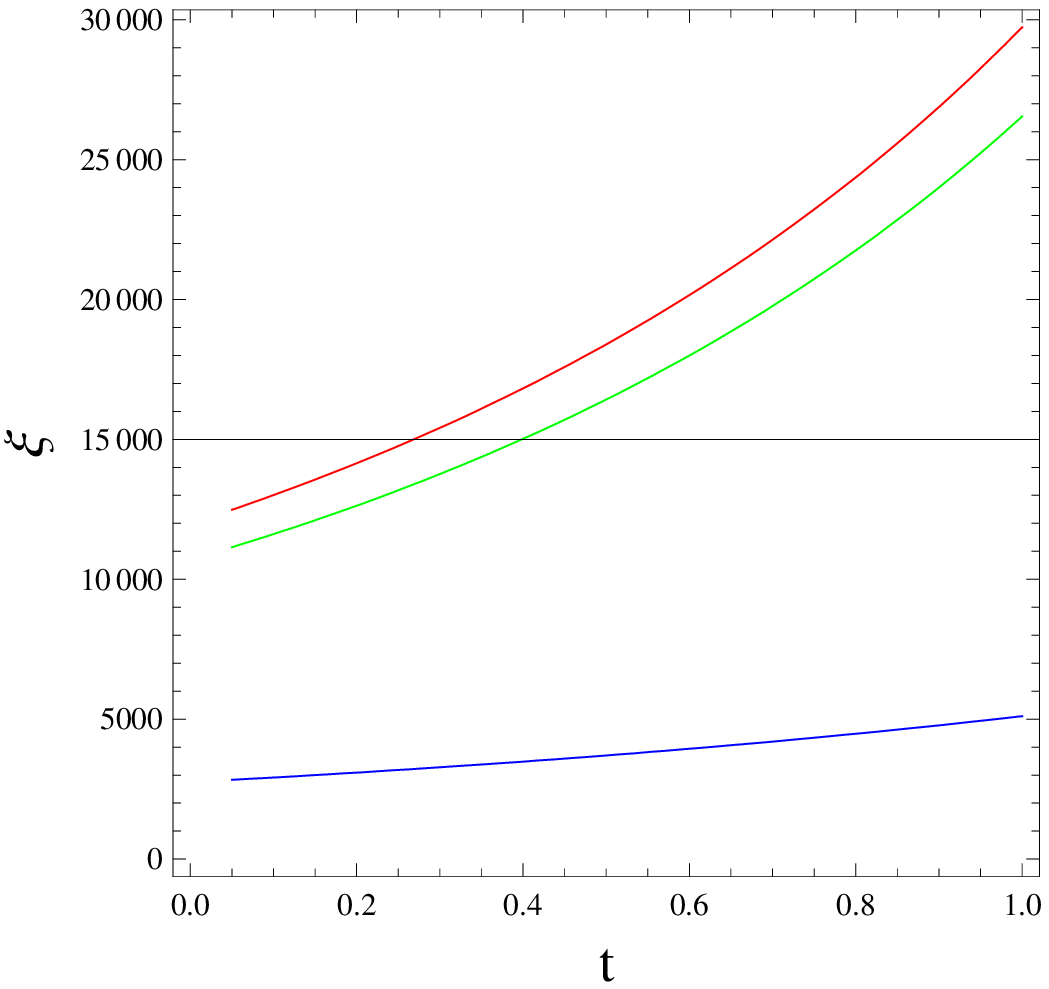}
\caption{\label{figxi}Plot of the bulk viscosity coefficient $\xi$. Red, green and blue lines correspond to $\{\alpha,\beta\}$ combination of $\{0.9735,3701\}$ with $\eta=3.5\times 10^{-4},3.0\times 10^{-4}$ and $2.5\times 10^{-4}$ respectively.}
\end{figure}
In Fig.\ref{figEoS} we have plotted the EoS parameter and observed that $w_{DE}<-1$ i.e. the phantom phase is attained. It is noted that the phantom barrier $w_{DE}=-1$ is never being crossed and in the very late stage $w_{DE}\ll -1$. In Fig. \ref{figpeff} the effect of bulk viscosity on the thermodynamic pressure is visualized. We observe that $p+\Pi\ll 0$ during the evolution. Fig.\ref{figpidot} shows that $\dot{\Pi}>0$ and from this we can understand that the effect of bulk viscosity on the thermodynamic pressure is increasing with evolution of the universe. However $\mid p_{eff}\mid$ is gradually decreasing with evolution of the universe. Fig.\ref{figxi} shows that the bulk viscosity coefficient $\xi$ is a monotone increasing function of cosmic time.

\subsection{Israel-Stewart approach for $a(t)=(t-t_0)^{\frac{\beta}{1-\alpha}}$ }
In the previous section instead of making any assumption on scale factor, we have derived solution for scale factor in Eq. (\ref{a}). In the present section the scale factor is chosen as
\begin{equation}\label{aplus}
a(t)=a_0 (t-t_0)^{\frac{\beta }{1-\alpha }}
\end{equation}
This leads to
\begin{eqnarray}
H=\frac{\beta }{(t-t_0)(1-\alpha)}\label{Hnew}\\
\dot{H}=-\frac{\beta }{(t-t_0)^2 (1-\alpha )}\label{Hdotnew}
\end{eqnarray}
 Based on Eqs. (\ref{Hnew}) and (\ref{Hdotnew}) the reconstructed density of EHRDE is
\begin{eqnarray}\label{appdark}
\rho_{DE}=\frac{3 \beta ^2 (2 \alpha-1) }{(t-t_0)^2 (1-\alpha )^2}
\end{eqnarray}
For $\rho>0$ one needs $\alpha>1/2$. Accordingly the bulk viscosity coefficient and relaxation times (Eq. (\ref{xi})) get reconstructed and hence evolution equation for
$\Pi$ in the framework of the truncated Israel-Stewart theory given by Eq. (\ref{Pi}) is solved to get
\begin{equation}
\begin{array}{c}
\Pi=C_2 e^{-\frac{(t-t_0)^3 (-1+\alpha )^2}{9 (-1+2 \alpha ) \beta ^2 \eta }}+\\
\frac{e^{-\frac{(t-t_0)^3 (-1+\alpha )^2}{9 (-1+2 \alpha ) \beta ^2 \eta }} (-1+2 \alpha ) \beta ^3 \left(-9 e^{\frac{(t-t_0)^3 (-1+\alpha )^2}{9 (-1+2 \alpha ) \beta ^2 \eta }}+3^{2/3} \left(-\frac{(t-t_0)^3 (-1+\alpha )^2}{(-1+2 \alpha ) \beta ^2 \eta }\right)^{2/3} \Gamma\left[\frac{1}{3},-\frac{(t-t_0)^3 (-1+\alpha )^2}{9 (-1+2 \alpha ) \beta ^2 \eta }\right]\right)}{2 (t-t_0)^2 (-1+\alpha )^3}\\
\end{array}\label{appPai}
\end{equation}
and hence
\begin{equation}
\begin{array}{c}
\dot{\Pi}=\frac{e^{-\frac{(t-t_0)^3 (-1+\alpha )^2}{9 (-1+2 \alpha ) \beta ^2 \eta }} }{6 (t-t_0)^3 (-1+\alpha )^3 (-1+2 \alpha ) \beta ^2 \eta ^2}\times\\
\left(\eta  \left(-2 C_2 (t-t_0)^5 (-1+\alpha )^5+9 e^{\frac{(t-t_0)^3 (-1+\alpha )^2}{9 (-1+2 \alpha ) \beta ^2 \eta }} (-1+2 \alpha ) \beta ^3 \left((t-t_0)^3 (-1+\alpha )^2+6 (-1+2 \alpha ) \beta ^2 \eta \right)\right)+\right.\\
\left.(t-t_0)^6 (-1+\alpha )^4 \beta  E_{\frac{2}{3}}\left[T\right]\right)
\end{array}\label{appPaidot}
\end{equation}
where, $T=-\frac{(t-t_0)^3 (-1+\alpha )^2}{9(-1+2 \alpha ) \beta ^2 \eta }$. Subsequently, $p_{DE}$ becomes
\begin{equation}
\begin{array}{c}
p_{DE}=C_2 e^{-\frac{(t-t_0)^3 (-1+\alpha )^2}{9 (-1+2 \alpha ) \beta ^2 \eta }}-\frac{2 \beta }{(t-t_0)^2 (-1+\alpha )}\\
-\frac{3 \beta ^2}{(t-t_0)^2 (-1+\alpha )^2}+\frac{e^{-\frac{(t-t_0)^3 (-1+\alpha )^2}{9 (-1+2 \alpha ) \beta ^2 \eta }} (-1+2 \alpha ) \beta ^3 \left(9 e^{\frac{(t-t_0)^3 (-1+\alpha )^2}{9 (-1+2 \alpha ) \beta ^2 \eta }}-3^{2/3} \left(-\frac{(t-t_0)^3 (-1+\alpha )^2}{(-1+2 \alpha ) \beta ^2 \eta }\right)^{2/3} \Gamma\left[\frac{1}{3},-\frac{(t-t_0)^3 (-1+\alpha )^2}{9 (-1+2 \alpha ) \beta ^2 \eta }\right]\right)}{2 (t-t_0)^2 (-1+\alpha )^3}\\
\end{array}\label{apppde}
\end{equation}

and hence EoS parameter $w_{DE}$ is
\begin{equation}
\begin{array}{c}
w_{DE}=\frac{(t-t_0)^2 (-1+\alpha )^2 }{3 (-1+2 \alpha ) \beta ^2}\times\\
\left(-C_2 e^{-\frac{(t-t_0)^3 (-1+\alpha )^2}{9 (-1+2 \alpha ) \beta ^2 \eta }}-\frac{2 \beta }{(t-t_0)^2 (-1+\alpha )}-\right.\\
\left.\frac{3 \beta ^2}{(t-t_0)^2 (-1+\alpha )^2}+\frac{e^{-\frac{(t-t_0)^3 (-1+\alpha )^2}{9 (-1+2 \alpha ) \beta ^2 \eta }} (-1+2 \alpha ) \beta ^3 \left(9 e^{\frac{(t-t_0)^3 (-1+\alpha )^2}{9 (-1+2 \alpha ) \beta ^2 \eta }}-3^{2/3} \left(-\frac{(t-t_0)^3 (-1+\alpha )^2}{(-1+2 \alpha ) \beta ^2 \eta }\right)^{2/3} \Gamma\left[\frac{1}{3},-\frac{(t-t_0)^3 (-1+\alpha )^2}{9 (-1+2 \alpha ) \beta ^2 \eta }\right]\right)}{2 (t-t_0)^2 (-1+\alpha )^3}\right)
\end{array}\label{wapp}
\end{equation}

In the above equations $E_n[z]=\int_1^\infty \frac{e^{-zt}}{t^n}dt$.
\begin{figure}[ht]
 \begin{minipage}[b]{0.45\linewidth}
\centering\includegraphics[width=\textwidth]{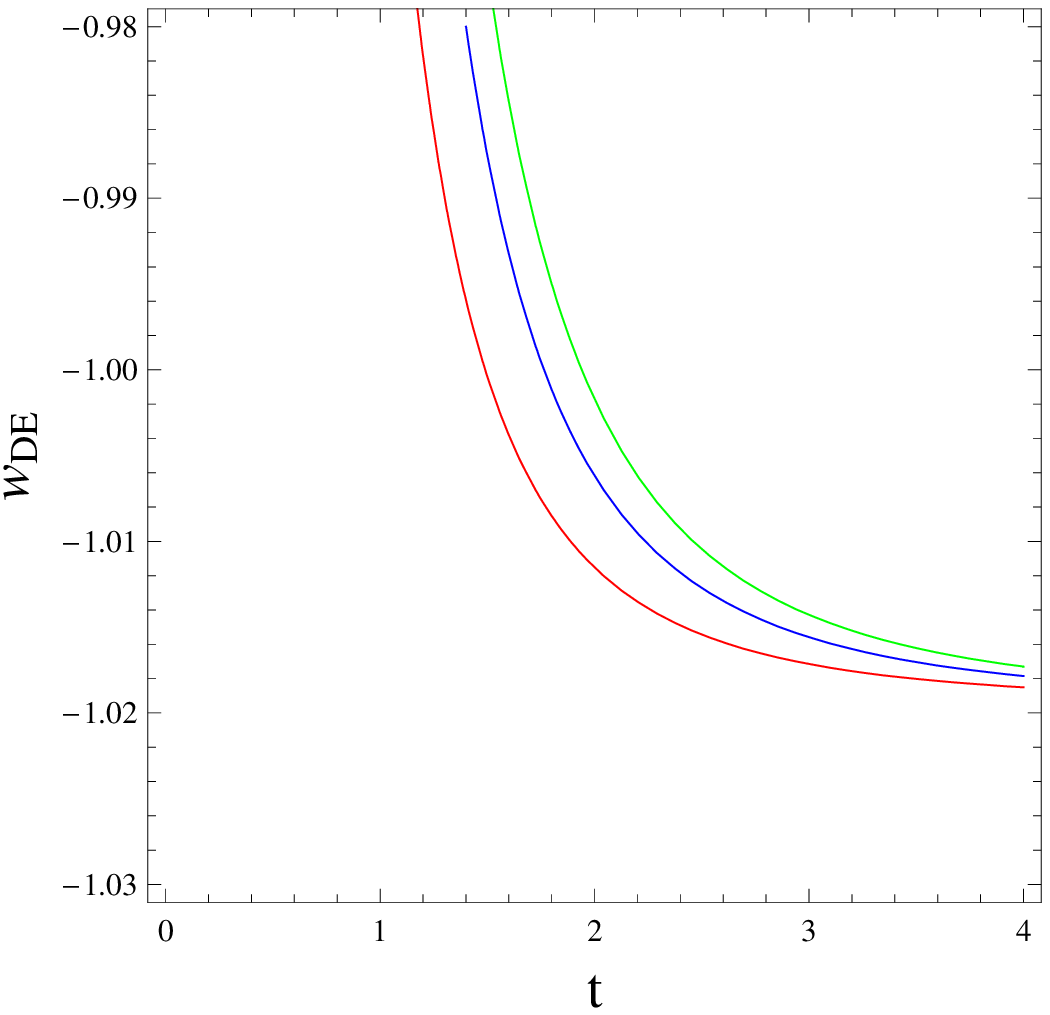} \caption{Plot of $w_{DE}$ for scale factor in Eq. (\ref{wapp})} \label{figapp} \end{minipage}
\hspace{0.5cm} \begin{minipage}[b]{0.45\linewidth}
\centering\includegraphics[width=\textwidth]{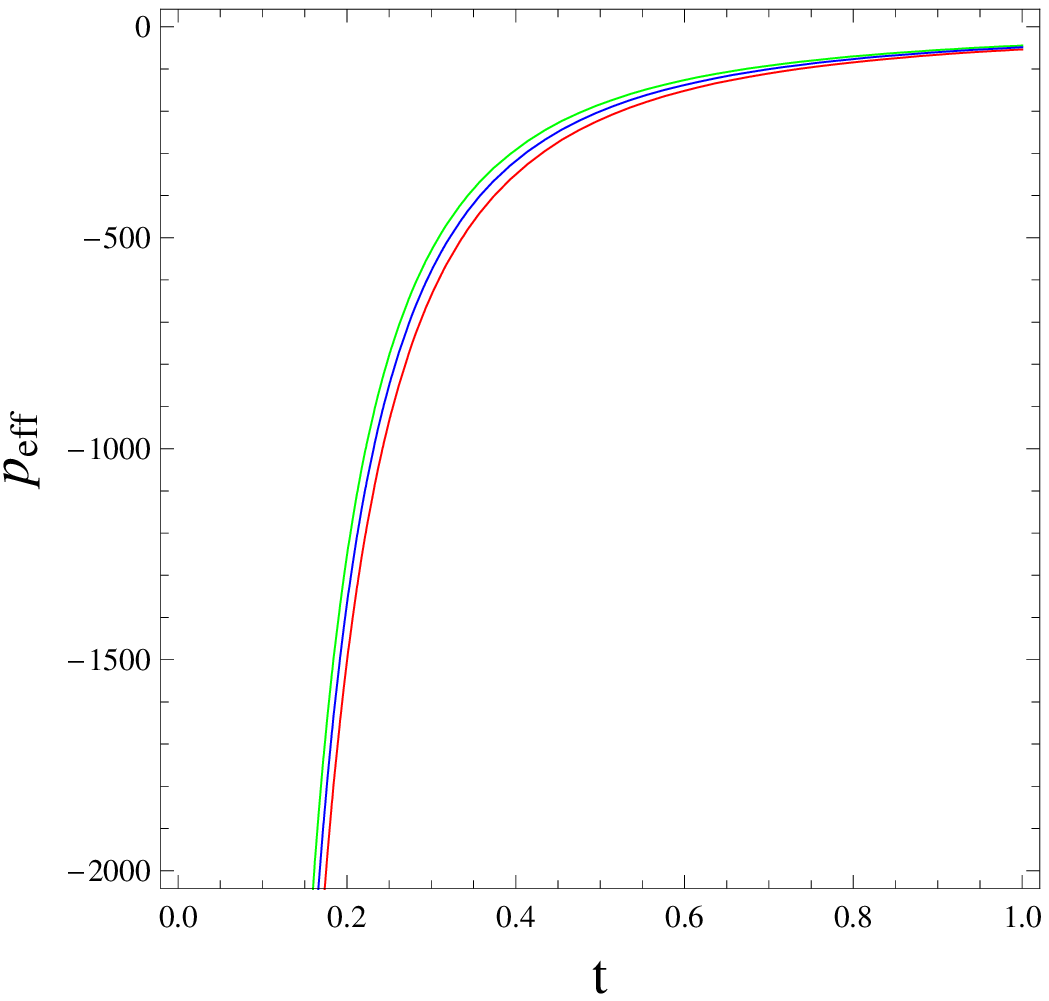}
\caption{Effective pressure $p_{eff}=p_{DE}+\Pi$ based on Eqs.(\ref{appPai}) and (\ref{apppde}).} \label{figappeff} \end{minipage}
\hspace{0.5cm}
\begin{minipage}[b]{0.45\linewidth}
\centering\includegraphics[width=\textwidth]{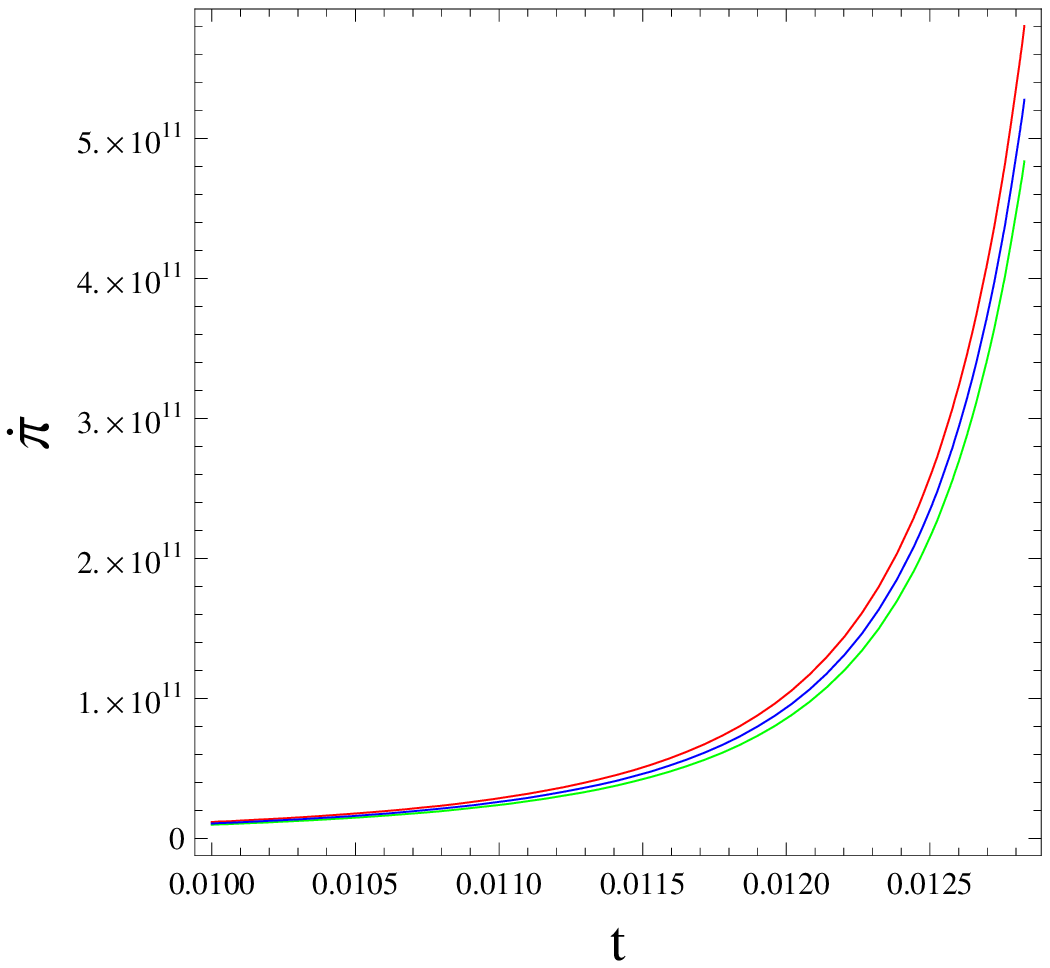}\caption{Time evolution of bulk viscous pressure $\dot{\Pi}$ based on Eq. (\ref{appPaidot}).}\label{figappPaidot}
\end{minipage}
\end{figure}

In Figs. \ref{figapp}, \ref{figappeff} and \ref{figappPaidot} the red, green and blue lines correspond to $\eta=0.00009,~0.0002$ and $0.00015$ respectively. In all the cases in this section $\alpha=0.9180,~\beta=0.3701$. In Fig. \ref{figapp} we plot the $w_{DE}$ and observe that the EoS parameter shows a clear transition feom $w_{DE}>-1$ to $w_{DE}<-1$ i.e. from quintessence to phantom. Hence, the model behaves like ``quintom". This is in contradiction with what happened in the model without any specific choice of scale factor, where the $w_{DE}<-1$ that implies ``phantom" behaviour of the EoS parameter. Similar to the earlier case, the effective pressure (Fig. \ref{figappeff}) is deacying with the evolution of the universe. However, time derivative of the bulk viscous pressure $\dot{\Pi}$ stays at positive level (Fig. \ref{figappPaidot}). This indicates that the effect of bulk viscous pressure increases with time. This means that the negative bulk viscous pressure gives a significant contribution to the total negative pressure of the EHRDE. The positive time derivative of bulk viscous pressure and gradually decaying effective pressure indicates that non-equilibrium bulk viscous pressure is small compared to the local equilibrium pressure.

\section{Statefinder parameters}
\subsection{Model without any specific choice of scale factor}
Sahni et al. \cite{state1} and and Alam et al. \cite{state2} introduced a pair of cosmological parameters $\{r,s\}$ (the so-called
``statefinder parameters") that seem to be promising candidates for the purpose of discrimination between the various contenders of dark energy. If the $\{r-s\}$ trajectory meets the point $\{r=1,s=0\}$ then the model is said to attain $\Lambda$CDM phase of the universe. In the literature, quite a good number of works are available, where dark energy models have been explored through statefinder trajectories e.g. \cite{state3,state4,state5}. The statefinder parameters are given by
\begin{eqnarray}
r &=& q+2q^2+\frac{\dot{q}}{H}\label{r1} \\
s &=& \frac{r-1}{3\left(q-\frac{1}{2}\right)}. \label{s1}
\end{eqnarray}
In Eqs.(\ref{r1}) and (\ref{s1}) deceleration parameter $q$ is given by
\begin{equation}
q=-\frac{\ddot{a}a}{\dot{a}^2}
\end{equation}
where $a$ is the scale factor as available in Eq.(\ref{a}). Hence, in the current framework Eq.(\ref{r1}) and (\ref{s1}) take the form
\begin{figure}[h]
\includegraphics[width=20pc]{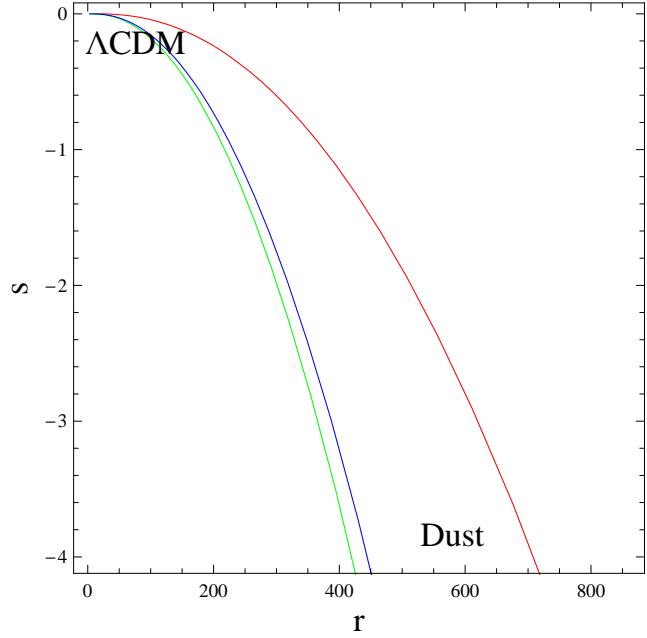}
\caption{\label{figrs}The $\{r-s\}$ trajectory based on Eqs.(\ref{r}) and (\ref{s}). Red, green and blue lines correspond to $\{\alpha,\beta\}$ combination of $\{0.9735,3701\}$ with $\eta=3.5\times 10^{-4},3.0\times 10^{-4}$ and $2.5\times 10^{-4}$ respectively.}
\end{figure}
\begin{equation}\label{r}
r=\frac{H_0^2 (\alpha-1 )^2 \left(-2+\frac{\beta }{\alpha-1 }\right) \left(-1+\frac{\beta }{\alpha-1 }\right)}{(H_0
(t-t_0) (\alpha-1 )+\beta )^2}
\end{equation}
\begin{equation}
\begin{array}{c}\label{s}
s=2 \left.\left(-1+\frac{H_0^2 (\alpha-1 -\beta ) (-2+2 \alpha -\beta )}{(H_0 (t-t_0) (\alpha-1 )+\beta )^2}\right)\right/\\
3\left(-3-\frac{2}{\beta }+\frac{2 \alpha }{\beta }+
\left(C_1 e^{-\frac{(H_0 (-1+t t_0) (\alpha-1 )-t \beta )^3}{9 H_0^2 (H_0 t_0 (\alpha-1 )-\beta ) \beta ^2 \eta
}}\right.\right. \\
\left.(H_0 (t-t_0) (\alpha-1 )+\beta )^2 \left(2 H_0^2 \left(1-4 t t_0+3 t^2 t_0^2\right) (\alpha-1 )^2 \beta -2 H_0
t (-2+3 t t_0) (\alpha-1 ) \beta ^2+2 t^2 \beta ^3+\right.\right.\\
\left.\left.\left.H_0^3 \left(-2 t_0 (\alpha-1 )^3+\right.\right.\right.\right.\\
\left.\left.\left.\left.4 t t_0^2 (\alpha-1 )^3-2 t^2 t_0^3 (\alpha-1 )^3+9 e^{\frac{(H_0
(-1+t t_0) (\alpha-1 )-t \beta )^3}{9 H_0^2 (H_0 t_0 (\alpha-1 )-\beta ) \beta ^2 \eta }} \beta ^3\right)-\right.\right.\right.\\
\left.\left.\left.3^{2/3} H_0^3
\beta ^3 \left(-\frac{(H_0 (-1+t t_0) (\alpha-1 )-t \beta )^3}{H_0^2 (H_0 t_0 (\alpha-1 )-\beta ) \beta ^2 \eta }\right)^{2/3}
\Gamma\left[\frac{1}{3},-\frac{(H_0 (-1+t t_0) (\alpha-1 )-t \beta )^3}{9 H_0^2 (H_0 t_0 (\alpha-1 )-\beta
) \beta ^2 \eta }\right]\right)\right)\right/\\
\left.\left.\left(2 H_0^2 (H_0 t_0 (\alpha-1 )-\beta ) \beta ^2 (H_0 (-1+t t_0) (\alpha-1 )-t \beta )^2\right)\right)\right)
\end{array}
\end{equation}
In Fig.\ref{figrs} we observe that the $\{r-s\}$ trajectory can attain the $\Lambda$CDM point i.e. $\{r=1,s=0\}$. Again, for finite $r$, we observe that $s\rightarrow -\infty$ that corresponds to dust phase. Thus, we may conclude that the viscous EHRDE interpolates between dust and $\Lambda$CDM phases of the universe. However, the model deviates significantly from $\Lambda$CDM.

\subsection{Model with scale factor $a(t)=(t-t_0)^{\frac{\beta}{1-\alpha}}$}
For the choice of scale factor $a(t)=(t-t_0)^{\frac{\beta}{1-\alpha}}$ we get from Eqs.(\ref{r1}) and (\ref{s1}) the statefinder parameters as
\begin{equation}
\begin{array}{c}
r=\frac{1}{2} \left(1+e^{-\frac{(t-t_0)^3 (-1+\alpha )^2}{9 (-1+2 \alpha ) \beta ^2 \eta }}\times\right.\\
\left.\frac{ \left(-2 C_2 (t-t_0)^2
(-1+\alpha )^3 \eta +e^{\frac{(t-t_0)^3 (-1+\alpha )^2}{9 (-1+2 \alpha ) \beta ^2 \eta }} \beta  \left(-4 (-1+\alpha )^2-6 (-1+\alpha ) \beta
+9 (-1+2 \alpha ) \beta ^2\right) \eta +(t-t_0)^3 (-1+\alpha )^2 \beta  E_{\frac{2}{3}}\left[T\right]\right)}{2 (1+\alpha  (-3+2 \alpha )) \beta ^2 \eta }\right.\\
\left.-\frac{1}{6 (1+\alpha  (-3+2 \alpha )) \beta ^5
\eta ^2}e^{-\frac{(t-t_0)^3 (-1+\alpha )^2}{9 (-1+2 \alpha ) \beta ^2 \eta }} (t-t_0)^2 (-1+\alpha )^3\right.\\
 \left.\left(\beta ^2 \eta  \left(\frac{e^{\frac{(t-t_0)^3
(-1+\alpha )^2}{9 (-1+2 \alpha ) \beta ^2 \eta }} (t-t_0) \left(-4 (-1+\alpha )^2
-6 (-1+\alpha ) \beta +9 (-1+2 \alpha ) \beta ^2\right)}{(-1+2
\alpha ) \beta }-\right.\right.\right.\\
\left.\left.\left.12 C_2 (-1+\alpha ) \eta +\frac{(t-t_0)^4 (-1+\alpha )^2 E_{-\frac{1}{3}}\left[T\right]}{(-1+2 \alpha ) \beta  \eta }+9 (t-t_0) \beta  E_{\frac{2}{3}}\left[T\right]\right)-\right.\right.\\
\left.\left.\frac{(t-t_0) \left(-2 C_2 (t-t_0)^2 (-1+\alpha )^3 \eta +
e^{\frac{(t-t_0)^3
(-1+\alpha )^2}{9 (-1+2 \alpha ) \beta ^2 \eta }} \beta  \left(-4 (-1+\alpha )^2-6 (-1+\alpha ) \beta +9 (-1+2 \alpha ) \beta ^2\right) \eta +(t-t_0)^3
(-1+\alpha )^2 \beta  E_{\frac{2}{3}}\left[T\right]\right)}{-1+2
\alpha }\right)+\right.\\
\left.\left(1+e^{-\frac{(t-t_0)^3 (-1+\alpha )^2}{9 (-1+2 \alpha ) \beta ^2 \eta }}\times\right.\right.\\
\left.\left.\frac{ \left(-2 C_2 (t-t_0)^2 (-1+\alpha
)^3 \eta +e^{\frac{(t-t_0)^3 (-1+\alpha )^2}{9 (-1+2 \alpha ) \beta ^2 \eta }} \beta  \left(-4 (-1+\alpha )^2-6 (-1+\alpha ) \beta +9 (-1+2
\alpha ) \beta ^2\right) \eta +(t-t_0)^3 (-1+\alpha )^2 \beta  E_{\frac{2}{3}}\left[T\right]\right)}{2 (1+\alpha  (-3+2 \alpha )) \beta ^2 \eta }\right)^2\right)\\
\end{array}\label{appr}
\end{equation}
and
\begin{figure}
\centering
\includegraphics[width=20pc]{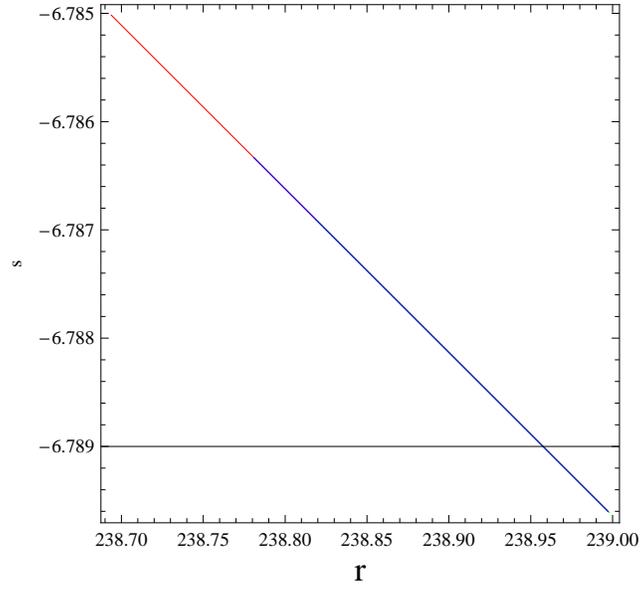}\\
\caption{Statefinder trajectories based on Eqs. (\ref{appr}) and (\ref{apps}).}\label{figappstate}
\end{figure}

\begin{equation}
\begin{array}{c}
s=\left(2 e^{\frac{(t-t_0)^3 (-1+\alpha )^2}{9 (-1+2 \alpha ) \beta ^2 \eta }}(1+\alpha  (-3+2 \alpha )) \beta ^2 \eta \left(-1+e^{-\frac{(t-t_0)^3
(-1+\alpha )^2}{9 (-1+2 \alpha ) \beta ^2 \eta }}\times\right.\right.\\
\left.\left.  \frac{ \left(-2 C_2 (t-t_0)^2 (-1+\alpha )^3 \eta +e^{\frac{(t-t_0)^3 (-1+\alpha )^2}{9
(-1+2 \alpha ) \beta ^2 \eta }} \beta  \left(-4 (-1+\alpha )^2-6 (-1+\alpha ) \beta +9 (-1+2 \alpha ) \beta ^2\right) \eta +(t-t_0)^3 (-1+\alpha
)^2 \beta  E_{\frac{2}{3}}\left[T\right]\right)}{2 (1+\alpha
 (-3+2 \alpha )) \beta ^2 \eta }\right.\right.\\
 \left.\left.-\frac{e^{-\frac{(t-t_0)^3 (-1+\alpha )^2}{9 (-1+2 \alpha
) \beta ^2 \eta }}}{6 (1+\alpha  (-3+2 \alpha )) \beta ^5 \eta ^2} (t-t_0)^2 (-1+\alpha )^3\times \right.\right.\\
 \left.\left.\left(\beta ^2 \eta  \left(\frac{e^{\frac{(t-t_0)^3 (-1+\alpha )^2}{9 (-1+2 \alpha ) \beta
^2 \eta }} (t-t_0) \left(-4 (-1+\alpha )^2-6 (-1+\alpha ) \beta +9 (-1+2 \alpha ) \beta ^2\right)}{(-1+2 \alpha ) \beta }-\right.\right.\right.\right.\\
\left.\left.\left.\left.12 C_2 (-1+\alpha
) \eta +\frac{(t-t_0)^4 (-1+\alpha )^2 E_{-\frac{1}{3}}\left[T\right]}{(-1+2 \alpha ) \beta  \eta }+9 (t-t_0) \beta  E_{\frac{2}{3}}\left[T\right]\right)-\right.\right.\right.\\
\left.\left.\left.\frac{(t-t_0) \left(-2 C_2 (t-t_0)^2 (-1+\alpha )^3 \eta +e^{\frac{(t-t_0)^3
(-1+\alpha )^2}{9 (-1+2 \alpha ) \beta ^2 \eta }} \beta  \left(-4 (-1+\alpha )^2-6 (-1+\alpha ) \beta +9 (-1+2 \alpha ) \beta ^2\right) \eta +(t-t_0)^3
(-1+\alpha )^2 \beta  E_{\frac{2}{3}}\left[T\right]\right)}{-1+2
\alpha }\right)+\right.\right.\\
\left.\left.\left(1+e^{-\frac{(t-t_0)^3 (-1+\alpha )^2}{9 (-1+2 \alpha ) \beta ^2 \eta }}\times\right.\right.\right.\\
\left.\left.\left.\frac{ \left(-2 C_2 (t-t_0)^2 (-1+\alpha
)^3 \eta +e^{\frac{(t-t_0)^3 (-1+\alpha )^2}{9 (-1+2 \alpha ) \beta ^2 \eta }} \beta  \left(-4 (-1+\alpha )^2-6 (-1+\alpha ) \beta +9 (-1+2
\alpha ) \beta ^2\right) \eta +(t-t_0)^3 (-1+\alpha )^2 \beta  E_{\frac{2}{3}}\left[T\right]\right)}{2 (1+\alpha  (-3+2 \alpha )) \beta ^2 \eta }\right)^2\right)\right)/\\
\left(3 \left(-2 C_2 (t-t_0)^2
(-1+\alpha )^3 \eta +\right.\right.\\
\left.\left.e^{\frac{(t-t_0)^3 (-1+\alpha )^2}{9 (-1+2 \alpha ) \beta ^2 \eta }} \beta  \left(-4 (-1+\alpha )^2-6 (-1+\alpha ) \beta
+9 (-1+2 \alpha ) \beta ^2\right) \eta +(t-t_0)^3 (-1+\alpha )^2 \beta  E_{\frac{2}{3}}\left[T\right]\right)\right)
\end{array}\label{apps}
\end{equation}
In Fig. \ref{figappstate} we have plotted the statefinder trajectories for the model with scale factor $a(t)=(t-t_0)^{\frac{\beta}{1-\alpha}}$. It is observed that the fixed $\Lambda$CDM point $\{r=1,s=0\}$ is not attained by the trajectories. However, for finite $r$, we observe $s\rightarrow -\infty$, that corresponds to the dust phase of the universe.

\section{Generalized second law of thermodynamics}

Discovery of black hole thermodynamics in 70's \cite{bek1,bek2,bek3} prompted physicists to study the thermodynamics of cosmological models of the universe. Semi classical description in black hole physics shows that a black hole behaves like a black body that is emitting thermal radiation with temperature and entropy. This temperature and the entropy are known as Hawking temperature and Bekenstein entropy respectively \cite{sir1,gs1,gs2}. This entropy is proportional to surface area $A$ of black hole, which, according to Hawking’s area theorem cannot decrease. Based on the conjectured proportionality between entropy and horizon area of black hole, a generalized version of the second law of thermodynamics was proposed by Bekenstein. According to the proposal of Bekenstein the sum of black hole entropy and the entropy of matter and radiation in the region exterior to black hole can not decrease. The GSL provides a relation between gravitation, thermodynamics and quantum theory. In a  very recent work \cite{busso} conjectured a novel GSL that can be applied in cosmology irrespective of the presence of event horizon.

Ref. \cite{setare2} examined the validity of the generalized GSL in a non-flat universe in the presence of viscous dark energy. Setare \cite{setarethermo} investigated the validity of the generalized second law of thermodynamics for the quintom model of dark energy. Considering the universe as a closed bounded system filled with $n$ component fluids Bamba et al. \cite{thermobamba} studied the generalized second law in $f(T)$ cosmology. Ref \cite{thermoherera} investigated the validity of the generalized second law in the context of interacting $f(R)$ gravity. We consider the universe
To check the generalized second law of thermodynamics, we have to examine the evolution of the
total entropy $S_A+S_{DE}$, where $S_A$ denotes the entropy of the apparent horizon and $S_{DE}$ denotes the entropy of the fluid inside the horizon. For the FRW universe the apparent horizon radius reads \cite{setare2}
\begin{equation}\label{rAk}
\tilde{r}_A=\frac{1}{\sqrt{H^2+\frac{k}{a^2}}}
\end{equation}
In a flat universe $k=0$ and Eq.(\ref{rAk}) becomes
\begin{equation}\label{rA}
\tilde{r}_A=\frac{1}{H}
\end{equation}
Temperature on the apparent horizon is defined as \cite{setare2}
\begin{equation}
T_A=\frac{1}{2\pi \tilde{r}_A}\left(1-\frac{\dot{\tilde{r}}_A}{2H\tilde{r}_A}\right)
\end{equation}
The entropy associated to the apparent horizon is \cite{setare2}
\begin{equation}
S_A=\frac{A}{4G}=\frac{\pi\tilde{ r}_A^2}{G}
\end{equation}
where where $A=4\pi \tilde{r}_A^2$ is the area of the apparent horizon. It has been shown by some calculations in \cite{setare2} that for viscous dark energy dominated flat universe enveloped by the apparent horizon
\begin{equation}
T_A\dot{S}_A=4\pi H \tilde{r}_A^3 \left(\rho_{DE}(1+w_{DE})-3H\xi\right)\left(1-\frac{\dot{\tilde{r}}_A}{2H\tilde{r}_A}\right)
\end{equation}
In the present case the above can be rewritten as
\begin{equation}\label{TASA}
T_A\dot{S}_A=4\pi H \tilde{r}_A^3 \left(\rho_{DE}(1+w_{DE})+\tau \dot{\Pi}+\Pi\right)\left(1-\frac{\dot{\tilde{r}}_A}{2H\tilde{r}_A}\right)
\end{equation}
The entropy of the viscous dark energy inside the apparent horizon, $S_{DE}$, can be related to its energy $E_D=\rho_D V$ and its pressure as
\begin{equation}\label{TDSD}
T_D\dot{S}_D=V\dot{\rho}_{DE}+(\rho_{DE}(1+w_{DE})-3H\xi)\dot{V}=V\dot{\rho}_{DE}+(\rho_{DE}(1+w_{DE})+\tau \dot{\Pi}+\Pi)\dot{V}
\end{equation}
where $T_{DE}$ and is the temperature of the viscous dark energy and $V=\frac{4}{3}\pi \tilde{r}_A^3$ is the volume enveloped
by the apparent horizon. Under the assumption that the thermal system bounded by the apparent horizon remains in equilibrium i.e. temperature of the system is uniform and the temperature on the horizon is equal to temperature of the fluid inside the horizon we have $T_A=T_{DE}=T$. It has been shown in \cite{setare2} by some simple calculation that
\begin{equation}\label{Sdottotal}
T\left(\dot{S}_{A}+\dot{S}_{DE}\right)=\frac{A}{2}\left(\rho_{DE}(1+w_{DE})+\tau \dot{\Pi}+\Pi\right)\dot{\tilde{r}}_A
\end{equation}
\begin{figure}[h]
\includegraphics[width=25pc]{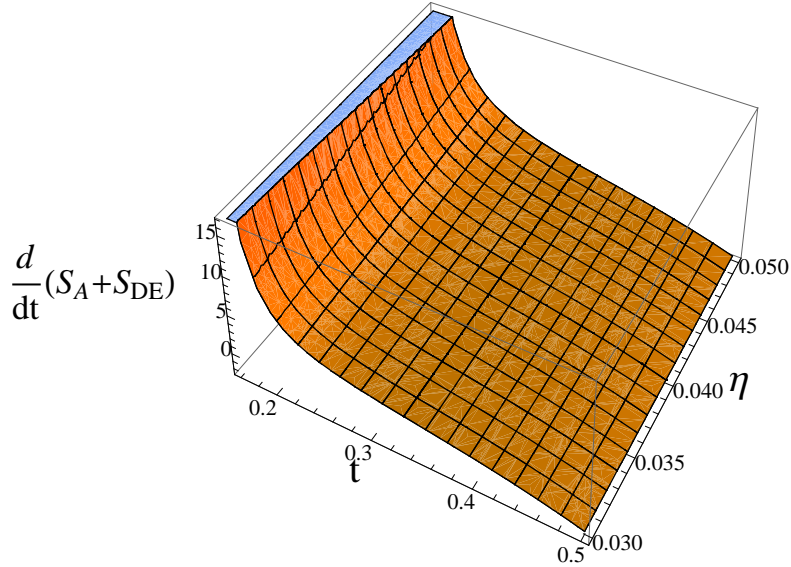}
\caption{\label{figGSL}Plot of time evolution of total entropy $(\dot{S}_A+\dot{S}_{DE})$ based on Eq.(\ref{TSdottotal}). The $\{\alpha,\beta\}$ combination is taken as $\{0.9735,0.3701\}$ and $\eta$ ranges from 0.0002 to 0.0005.}
\end{figure}
Using Eqs.(\ref{RDreconst}),(\ref{EoS}) and (\ref{Pisolve}) in Eq.(\ref{Sdottotal}) we get
\begin{equation}
\begin{array}{c}\label{TSdottotal}
\dot{S}_A+\dot{S}_{DE}=
\frac{2 \pi}{H_0 \beta ^3}  (-1+\alpha ) (H_0 (t-t_0) (-1+\alpha )+\beta ) \left(\frac{\beta }{G}-\frac{8 \pi}{\alpha
-2 \beta-1 }\times\right.\\
\left.  \left(-2 \beta ^2+\frac{1}{H_0^2}(H_0 (t-t_0) (-1+\alpha )+\beta )^2 \left(C_1 e^{-\chi }-\frac{e^{-\chi
} H_0^3 \beta ^3 \left(9 e^{\chi }-3^{2/3} (-\chi )^{2/3} \Gamma\left[\frac{1}{3},-\chi \right]\right)}{2 (H_0 t_0 (-1+\alpha
)-\beta ) (H_0 (-1+t t_0) (-1+\alpha )-t \beta )^2}+e^{-\chi } \beta ^2 \eta \times \right.\right.\right.\\
\left.\left.\left.\frac{
 \left(-\frac{2 C_1 (H_0 (-1+t t_0)
(-1+\alpha )-t \beta )^2}{\beta ^2 \eta }-\frac{6 3^{2/3} H_0^3 \beta  \Gamma\left[\frac{1}{3},-\chi \right]}{(H_0 t_0 (-1+\alpha
)-\beta ) \eta  (-\chi )^{1/3}}+\frac{6 H_0^5 \beta ^3 \left(9 e^{\chi }-3^{2/3} (-\chi )^{2/3} \Gamma\left[\frac{1}{3},-\chi \right]\right)}{(H_0
(-1+t t_0) (-1+\alpha )-t \beta )^3}+\frac{H_0^3 \beta  \left(9 e^{\chi }-3^{2/3} (-\chi )^{2/3} \Gamma\left[\frac{1}{3},-\chi
\right]\right)}{(H_0 t_0 (-1+\alpha )-\beta ) \eta }\right)}{2 (H_0 (t-t_0) (-1+\alpha )+\beta )^2}+\right.\right.\right.\\
\left.\left.\left.\frac{H_0^2 \left(2
\beta (\alpha-1) -\frac{e^{-\chi } (H_0 (t-t_0) (-1+\alpha )+\beta )^2 \left(-9 e^{\chi } H_0^3 \beta ^3+2 C_1 (H_0
t_0 (-1+\alpha )-\beta ) (H_0 (-1+t t_0) (-1+\alpha )-t \beta )^2+3^{2/3} H_0^3 \beta ^3 (-\chi )^{2/3} \Gamma\left[\frac{1}{3},-\chi
\right]\right)}{2 H_0^2 (H_0 t_0 (-1+\alpha )-\beta ) (H_0 (-1+t t_0) (-1+\alpha )-t \beta )^2}\right)}{(H_0
(t-t_0) (-1+\alpha )+\beta )^2}\right)\right)\right)
\end{array}
\end{equation}
where $\chi=\frac{(H_0 (-1+t t_0) (-1+\alpha )-t \beta )^3}{9 H_0^2 (H_0 t_0 (-1+\alpha )-\beta ) \beta ^2 \eta
}$. In Fig.\ref{figGSL} we have plotted $T(\dot{S}_A+\dot{S}_{DE})$ based on Eq. (\ref{TSdottotal}) with cosmic time $t$ and $\eta$ ranging from 0.0002 to 0.0005. We have observed that $T(\dot{S}_A+\dot{S}_{DE})$ is staying at positive level. Since $T>0$, Fig. \ref{figGSL} indicates that $\dot{S}_A+\dot{S}_{DE}\geq 0$.  This indicates validity of the generalized second law  of thermodynamics in an universe dominated by viscous EHRDE.
\begin{figure}[h]
\includegraphics[width=25pc]{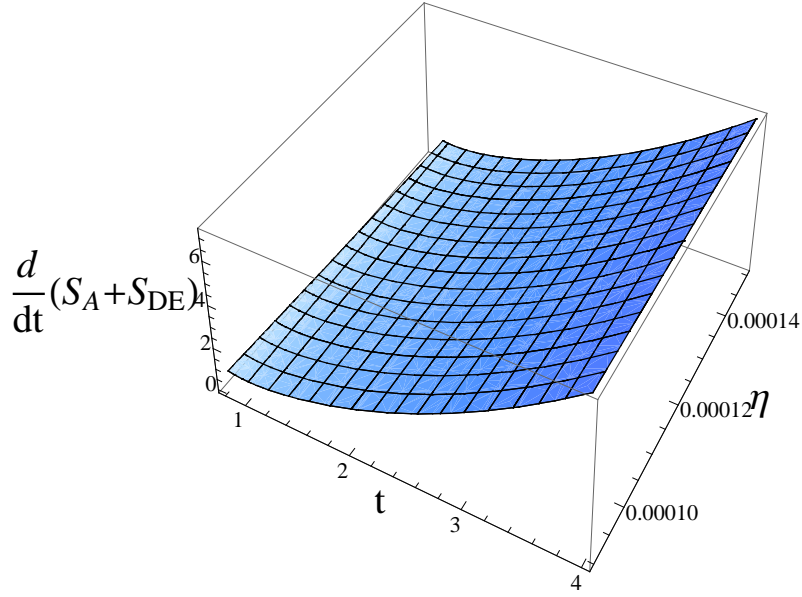}
\caption{\label{figappGSL}Plot of time evolution of total entropy $(\dot{S}_A+\dot{S}_{DE})$ based on Eq.(\ref{TSdottotal}) for $a(t)=(t-t_0)^{\frac{\beta}{1-\alpha}}$. The $\{\alpha,\beta\}$ combinations is taken as $\{0.9180,0.3701\}$ and $\eta$ ranges from 0.00009 to 0.00015.}
\end{figure}
Using Eqs.(\ref{appdark}),(\ref{appPai}) and (\ref{wapp}) in Eq.(\ref{Sdottotal}) we get the time evolution of the total entropy for the case $a(t)=(t-t_0)^{\frac{\beta}{1-\alpha}}$ as
\begin{equation}
\begin{array}{c}
\dot{S}_A+\dot{S}_{DE}=
\frac{1}{\beta ^2 (-1+\alpha +2 \beta )} 8 \pi ^2 (t-t_0) (-1+\alpha )^2 \\
\left(2+(-t+t_0) (-1+\alpha )-2 \alpha -\frac{C_2 e^{-\frac{(t-t_0)^3 (-1+\alpha )^2}{9 (-1+2 \alpha ) \beta ^2 \eta }} (t-t_0)^2
(-1+\alpha )^2}{\beta }-3 \beta +3 (-1+2 \alpha ) \beta +\right.\\
\left.\frac{9 (-1+2 \alpha ) \beta ^2}{2 (-1+\alpha )}+(-1+\alpha ) \left(\frac{3 (-1+2 \alpha ) \beta ^2 \eta }{(t-\text{t0})^2 (-1+\alpha )^2}\right) +\frac{e^{-\frac{(t-t_0)^3
(-1+\alpha )^2}{9 (-1+2 \alpha ) \beta ^2 \eta }} (t-t_0)^3 (-1+\alpha ) E_{\frac{2}{3}}\left[T\right]}{2 \eta }\right)
\end{array}
\end{equation}
Fig. \ref{figappGSL} shows that like the previous case $\dot{S_{total}}=\dot{S}_A+\dot{S}_{DE}>0$ and hence the GSL of thermodynamics is validated in an universe dominated by viscous EHRDE and expanding according to $a(t)=(t-t_0)^{\frac{\beta}{1-\alpha}}$. However, contrary to Fig. \ref{figGSL}, the time derivative of total entropy is increasing with evolution of the universe.
\section{Conclusions}
Motivated by \cite{vcg} and \cite{feng} we have presented a study on viscous extended holographic Ricci dark energy (EHRDE) in flat FRW universe based on Israel-Stewart approach. The work has been carried out in two phases. In one phase instead of choosing any specific form of scale factor we have reconstructed Hubble parameter $H$ based on the field equation (Eq. \ref{f1}) with $\rho=\rho_{DE}=3 M_p^2\left(\alpha H^2+\beta \dot{H}\right)$ and subsequently solving the truncated Israel-Stewart theory given by $\tau \dot{\Pi}+\Pi=-3H\xi$ for $\Pi$ we studied the behaviour of EoS parameter $w_{DE}$, bulk viscous pressure $\Pi$, statefinder parameters and the thermodynamic consequences in terms of generalized second law (GSL) of thermodynamics. In this phase of study we have observed the following:

Under the consideration that the universe is dominated by EHRDE we have taken evolution equation for the bulk viscous pressure $\Pi$ in the framework of the truncated Israel-Stewart theory as $\tau \dot{\Pi}+\Pi=-3\xi H$, where $\tau$ is the relaxation time and $\xi$ is the bulk viscosity coefficient. Considering effective pressure as a sum of thermodynamic pressure of EHRDE and bulk viscous pressure we have observed that under the influence of bulk viscosity the EoS parameter $w_{DE}$ is behaving like phantom i.e. $w_{DE}\leq -1$ (see Fig.\ref{figEoS}). Furthermore, it has been observed that the effect of bulk viscosity is not blowing up in the very late stage or very early stage of the universe. This observation is consistent with \cite{feng}. Some constraints have been derived for the model parameters $\alpha$ and $\beta$ and other constants. Obeying the observational studies that show that for EHRDE $\alpha=0.8502^{+0.0984+0.1299}_{-0.0875-0.1064}$ and $\beta=0.4817^{+0.0842+0.1176}_{-0.0773-0.0955}$ we have taken $\alpha=0.9735, ~\beta=0.3701$. It has been already mentioned that like \cite{vis2} to obtain solution with big-rip no restriction is imposed on $\nu$. Fig. \ref{figEoS} shows that the $w_{DE}$ is decaying with evolution of the universe. It is observable that along with $\rho+p<0$ we also have $\rho+3p<0$ and with passage of time. Also, with passage of cosmic time $-(\rho+3p)$ increases, and so all gravitationally bound systems will be dissociated \cite{phantom} and the universe will end up in Big Rip. We have observed that the magnitude of the effective pressure $p_{eff}=p+\Pi$ is decaying with time (see Fig.\ref{figpeff}). Moreover, the non-negative derivative of $\Pi$ (see Fig.\ref{figpidot}) has indicated that the effect of bulk viscosity is increasing with time. The bulk viscosity coefficient $\xi$ has been found to be a increasing function of time (see Fig.\ref{figxi}). The statefinder parameters $\{r,s\}$ have also been studied. It has been observed that under the effect of bulk viscosity the statefinder trajectory $\{r-s\}$ for EHRDE is capable of reaching the $\Lambda$CDM point i.e. $\{r=1,s=0\}$. We have further observed that for finite $r$, the $s\rightarrow -\infty$, which corresponds to dust phase. This shows that the viscous EHRDE interpolates between dust and $\Lambda$CDM phases of the universe (see Fig.\ref{figrs}). Finally we have studied the thermodynamics of the viscous EHRDE under the assumption that the universe is enveloped by apparent horizon. We have derived the expression of the time derivative of the total entropy and we have observed that (see Fig.\ref{figGSL}) the time derivative is positive throughout the evolution of the universe. This shows that the generalized second law of thermodynamics is valid for the viscous EHRDE.

In the next phase of the study we have chosen the scale factor as $a(t)=(t-t_0)^{\frac{\beta}{1-\alpha}}$ and subsequently studied the cosmological parameters similar to that in the previous phase. The effective pressure and time derivative of bulk viscous pressure $\Pi$ appear to be of similar pattern to that in the earlier phase. However, the significant difference is observed in the case of EoS parameter $w_{DE}$ (see Fig. \ref{figapp}) and the statefinder trajectories $\{r-s\}$ (see Fig. \ref{figappstate}). The EoS parameter $w_{DE}$ is found to cross the phantom barrier i.e. transiting from quintessence $(w_{DE}>-1)$ to phantom $(w_{DE}<-1)$.Thus, the model is found to behave like ``quintom" for the said form of scale factor. In the statefinder trajectory, unlike the orevious case, the $\Lambda$CDM fixed point is not attainable and the model is deviated significantly from $\Lambda$CDM. However, the dust phase is attainable like the previous case. Like the previous case the GSL is found to be valid i.e. time derivative of the total entropy stays in the non-negative level. However, contrary to what happened in the previous case, the time derivative of the total entropy is increasing with evolution of the universe.

\section{Acknowledgement}
Financial support from DST, Govt of India under project grant no. SR/FTP/PS-167/2011 is thankfully acknowledged. The author sincerely thanks the reviewers for constructive suggestions. Warm hospitality provided by the Inter-University Centre for Astronomy and Astrophysics (IUCCA), Pune, India during a scientific visit in May 2016 is duly acknowledged by the author.
\\
\section{Statement on conflict of interest}
The author declares that there is no conflict of interest regarding the publication of this article.

\end{document}